%% file: sample-sigconf.tex
\documentclass[sigconf]{acmart}
\usepackage{booktabs} % For formal tables
\usepackage{hyperref}
\usepackage{comment}
\usepackage{multirow}
\usepackage{tabularx}
\usepackage{rotating}
\usepackage[inline]{enumitem}

\setcopyright{none}

\begin{document}
\title{IceBreaker: Solving Cold Start Problem for Video Recommendation Engines }%Content-based Video Relevance Prediction Challenge}
%\titlenote{Produces the permission block, and copyright information}
%\subtitle{Lipper : A System for Speech Reconstruction from Silent Videos}
%\subtitlenote{The full version of the author's guide is available as \texttt{acmart.pdf} document}

\author{Agniv Sharma}
%\authornote{Equal Contribution}
\authornotemark[1]
%\authornote{Dr.~Trovato insisted his name be first.}
\orcid{1234-5678-9012}
\affiliation{%
  \institution{DTU} 
  %\streetaddress{Delhi}
  \city{New Delhi} 
}
\email{agnivsharma96@gmail.com}

\author{Yaman Kumar}
\authornotemark[1]
%\authornote{Equal Contribution}
%\authornote{Dr.~Trovato insisted his name be first.}
\orcid{1234-5678-9012}
\affiliation{%
  \institution{Adobe Systems}
  %\streetaddress{Delhi}
  \city{Noida}
}
\email{ykumar@adobe.com}

\author{Abhigyan Khaund}
\authornote{Equal Contribution}
\orcid{1234-5678-9012}
\affiliation{%
  \institution{IIT Mandi}
  %\streetaddress{Delhi}
  \city{Mandi}
}
\email{khaundabhigyan@gmail.com}

\author{Akash Kumar}
\authornotemark[1]
%\authornote{Dr.~Trovato insisted his name be first.}
\orcid{1234-5678-9012}
\affiliation{%
  \institution{DTU}
  %\streetaddress{Delhi}
  \city{New Delhi}
}
\email{akash_bt2k15@dtu.ac.in}

\author{Ponnurangam Kumaraguru}
%\authornote{Dr.~Trovato insisted his name be first.}
\orcid{1234-5678-9012}
\affiliation{%
  \institution{IIIT-Delhi}
  %\streetaddress{Delhi}
  \city{New Delhi}
}
\email{pk@iitd.ac.in}

\author{Rajiv Ratn Shah}
%\authornote{Dr.~Trovato insisted his name be first.}
\orcid{1234-5678-9012}
\affiliation{%
  \institution{IIIT-Delhi}
  %\streetaddress{Delhi}
  \city{New Delhi}
}
\email{rajivratn@iiitd.ac.in}

\renewcommand{\shortauthors}{Yaman Kumar et al.}

\begin{abstract}
Internet has brought about a tremendous increase in content of all forms and, in that, video content constitutes the major backbone of the total content being published as well as watched. Thus it becomes imperative for video recommendation engines such as Hulu to look for novel and innovative ways to recommend the \emph{newly} added videos to their users. However, the problem with new videos is that they lack any sort of metadata and user interaction so as to be able to rate the videos for the consumers. To this effect, this paper introduces the several techniques we develop for the Content Based Video Relevance Prediction (CBVRP) Challenge being hosted by Hulu for the ACM Multimedia Conference 2018. We employ different architectures on the CBVRP dataset to make use of the provided frame and video level features and generate predictions of videos that are similar to the other videos. We also implement several ensemble strategies to explore complementarity between both the types of provided features. The obtained results are encouraging and will impel the boundaries of research for multimedia based video recommendation systems.

\end{abstract}


%%%%%%%
%\ccsdesc[500]{Human-centered computing~HCI design and evaluation methods}
%\ccsdesc[500]{Computing methodologies~Reconstruction}
%\ccsdesc[500]{Computing methodologies~Neural networks}
%%%%%%%%%


\maketitle

\input{samplebody-conf}

\bibliographystyle{ACM-Reference-Format}
\bibliography{sample-bibliography}

\end{document}

%% file: samplebody-conf.tex
\section{Introduction}
The recent years have seen a huge surge in online content, and, content in the form of videos has led the way. This stupendous wave has been steered by online video recommendation systems like Hulu, Youtube, Netflix, \emph{etc.} 
At the same time, the content supplied by traditional media houses and other licensed broadcasters has seen a plunge \cite{cha2007tube} with respect to amateurs and non-professionals uploading home-made content on various platforms. The popularity of video streaming can be judged from the fact that sixty hours of video content is generated every minute on Youtube \cite{bullas_agius_kakkar_vasileva_totka_savannah}. A tremendous amount of \emph{new} content is being generated at a very fast pace and, on top of that, the content metadata itself ranges from incorrect to either none or highly limited information \cite{deldjoo2016content, davidson2010youtube}. Additionally, the content thus generated is short (lesser than 10 minutes) and un-reviewed. The user interaction with the horde of these videos is minimal at the very best \cite{davidson2010youtube}. Due to these reason, video recommendation systems suffer from a multi-dimensional challenge where they have to deal with un-audited, un-rated and completely new content of which they know nothing about. This leads to the infamous \emph{cold start} problem where recommendation engines have to tackle new content and then process and recommend it to the users in such a manner so as to keep grabbing the eyeballs and dissuade them from shifting to the competitor's platform. 

Video recommendation has traditionally been inspired from two types of data: implicit and explicit. While on one hand, video recommendation on the basis of implicit content deals with features like light, mood, color, shape, motion, plot and other aesthetics; on the other hand, explicit content incorporates movie genre, director, actors, description, previous views and likes. Researchers have spent considerable time and effort building recommendation engines based on both the types of data \cite{deldjoo2016content,zhu2013videotopic,basharat2008content,basu1998recommendation,cbvrp-acmmm-2018}. But as noted in the previous section, for today's content explosion, the recommendation on the basis of explicit content fails utterly. This leaves one with implicit features. According to Applied Media Aesthetic \cite{zettl2002essentials}, media features such as light, color, camera motion \emph{etc.} serve as crucial elements which render emotional, aesthetic and informative effects. Thus a \emph{new} artistic work, can be appropriately evaluated and recommended on the basis of intrinsic features.

In this paper, we present our extensive experiments in building a video recommendation system which deals with the cold-start problem and uses intrinsic features to recommend videos. Here, we portray our experiments and their results over multiple architectures for Content Based Video Relevance Prediction Challenge of ACM Multimedia 2018, sponsored by Hulu.

\section{Related Work}
Building recommendation engines in general, and video recommendation engines in particular has always been a hot topic of research. Right from the development of the first recommender system, Tapestry\cite{goldberg1992using}, recommender systems have undergone drastic developmental changes. There have been several works which have employed the low level features, also known as stylistic features for the task of video recommendation\cite{yang2007online,zhao2011integrating,canini2013affective,lehinevych2014discovering}. But none of these approaches use \emph{only} low-level features, thus they cannot solve the cold-start problem in its full glory\cite{rubens2015active,deldjoo2016content,shah2017multimodal}. However, a few authors have realized this problem and have tried to tackle this problem head-on using various methods \cite{deldjoo2016content,wei2017collaborative, shah2014advisor}. Thus one of the novelties introduced in this work is exploring the recommendation problem using \emph{only} stylistic and artistic features for solving the daunting cold-start dilemma. 

\section{Methodology}
\label{technical_approach}
In this section, we present the dataset and different architectures \emph{i.e.,} random forest regression on video pairs,
deep learning based regression and neural network with DeepLDA for the CBCRP challenge. 
\subsection{Dataset}
In this paper, we have used the dataset from the CBVRP challenge\cite{cbvrp-acmmm-2018}. It consists of two tracks of data: \begin{enumerate*}
\item TV shows and
\item Movies
\end{enumerate*}. Track one consists of pre-extracted features from 7536 TV shows trailers and has been divided into three sets - training (3000 shows), validation (864 shows), and testing (3000 shows). Track two consists of pre-extracted features from 10826 movie trailers and has been divided into the following sets -  training (4500 movies), validation (1188 movies), and testing (4500 movies). The final results were evaluated on the test set by the organizers.

In both the given tracks, the pre-extracted feature vectors were composed of frame-level and video-level features. For frame-level feature vectors, frames of the video at 1 fps were passed into the InceptionV3 network\cite{youtube-8m} trained on ImageNet dataset\cite{deng2009imagenet}, and output of the ReLU\cite{reLU} activation of the last hidden layer of dimension 2048 was taken as the frame level feature. For video level features, each video at 8 fps was passed into the C3D network\cite{tran2015learning} and the activation of the pool5 layer of dimension 512 was taken as the feature vector.
Here we use these pre-extracted features from the training dataset as the input to our models.

\begin{figure*}
\includegraphics[scale = 0.35]{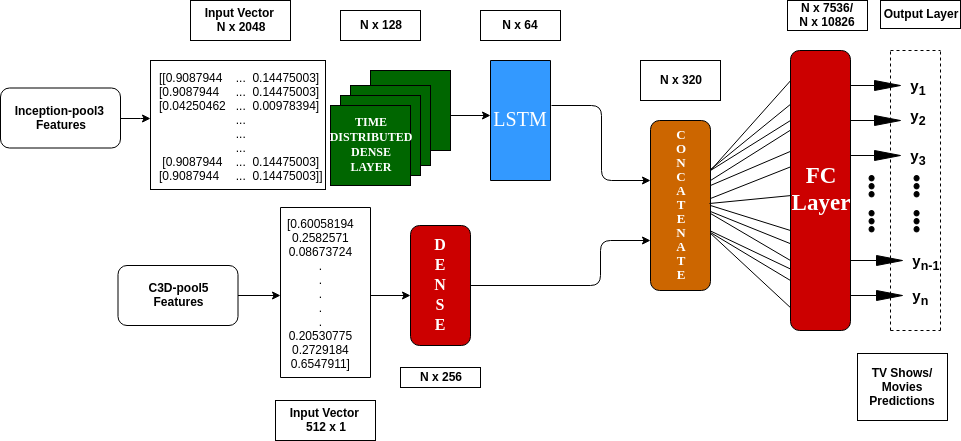}
\caption{Deep Learning based Regression}
\label{deep_learning_regression_model}
\end{figure*}

\subsection{Random Forest Regression on Video Pairs}
\begin{figure}
\includegraphics[width = 7cm, scale = 0.3]{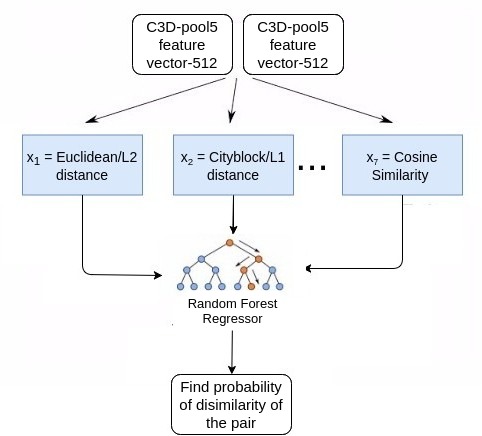}
\caption{Random Forest based regression on video pairs}
\label{rand_forest_model}
\end{figure}

\label{RandomForest Regression on pair}

%In order to find relation between videos, we make use of the video features from the C3D model and subsequently, find the closeness of these vectors using different distance metrics. This mechanism allows us to predict similar videos based on the local spatio-temporal features that are extracted from the C3D model.

To determine the closeness of two given C3D feature vectors, we consider several simplistic distance metrics - Euclidean\cite{deza2006dictionary}, Cityblock\cite{krause1975taxicab}, Chebyshev\cite{deza2006dictionary}, Correlation\cite{cha2007comprehensive,deza2006dictionary}, Square Euclidean, Braycurtis\cite{bray1957ordination,deza2006dictionary} and Cosine Similarity\cite{singhal2001modern,deza2006dictionary} .

\iffalse
For two vectors $\alpha$ and $\beta$ these distance metrics are defined as-

Euclidean/L2 distance: $\sqrt[]{\sum{(\alpha - \beta)^{2}}}$

Cityblock/L1 distance: $\sum{|\alpha - \beta|}$

Chebysev distance: $\max{|\alpha - \beta|}$ 

Correlation distance : $1 - \frac{(\alpha - \bar{\alpha})\cdot(\beta - \bar{\beta})}{||\alpha - \bar{\alpha}||_{2}||\beta - \bar{\beta}||_{2}}$ 

Squared Equilidean - $\sum{(\alpha - \beta)^{2}}$

Braycurtis distance - $\frac{\sum{|\alpha - \beta|}}{\sum{|\alpha + \beta|}}$

Cosine similarity - $1 - \frac{(\alpha)\cdot(\beta)}{||\alpha ||_{2}||\beta||_{2}}$
\fi
%In the ground truth of training data, for each video we have a set of other videos which are similar to that video. 

For a particular video, its positive pairs are formed with the videos that are in its ground truth. Negative pairs are formed by choosing videos randomly from the videos not present in the list of similar videos. To make the model not biased towards any type of pair\cite{deza2006dictionary}, we keep the number of positive pairs equal to the number of negative pairs. For each such pair, we make a vector $\textbf{X} =[x_{1},x_{2},x_{3},...x_{7}]$  where each $x_{i}$ is one of the distance types mentioned above calculated between the C3D vectors of each pair. The ground truth value corresponding to each pair is taken as $0$ for a positive pair and $1$ for a negative pair. We then use Random Forest \cite{liaw2002classification} to regress and choose the best features. 
%To be specific, the input to the random forest is an array of all the pairs which are formed from the training dataset. An overview of this architecture is presented in Figure \ref{rand_forest_model}. The expectation is that the random forest regressor would ensemble the result from different distance metrics and produce the best prediction on the similarity of the videos\cite{behnamian2017systematic}.

\subsection{Deep Learning based Regression}
\label{regression}
\subsubsection{Architecture Overview:} This model consists of two parallel networks as shown in Figure \ref{deep_learning_regression_model}. The first network consists of time distributed dense layers followed by LSTM layers and the second one contains only dense layers. The final layers concatenate the outputs of the two models and then output the target values for all the TV shows and Movies. The optimal number of layers in each network was found out through experimentation. We used the first network to process the frame-level features from the dataset. In the second network, we used video level features from the dataset. As the experiments panned out, creating a wider network rather than a deeper one proved to be more beneficial \cite{wu2016wider, zagoruyko2016wide, meghawat2018multimodal}. 

The size of the last layer is equal to the number of videos to be considered for that model. For a given video, it outputs the probability of the similarity of the other video to the given video. Subsequently, the top $K$ predictions are taken for the evaluation of results. In our model, we used the efficient ADAM\cite{adam} (lr=0.001) optimization algorithm and ReLU\cite{reLU} activation function\cite{chollet2015keras}. 
The same model was trained with losses as cosine proximity loss and poisson loss. The results for both the experiments are presented in Table \ref{results_shows}.

\iffalse
\subsubsection{Loss Functions:}
\paragraph{Cosine Proximity Loss: }
Cosine proximity loss function\cite{losses} computes the cosine distance between predicted value and the target value. The loss function equation is described as follows: 
\begin{equation}
\label{eq1}
\mathcal{L} = - \frac{\mathbf{x}.\mathbf{\hat{x}}}{||{\mathbf{x}}||_{2}.||\mathbf{\hat{x}}||_{2}} = - \frac{\sum_{i=1}^{n} \textit{x}^{(i)}. \textit{\^{x}}^{(i)}}{\sqrt{\sum_{i=1}^{n}(\textit{x}^{(i)})^{2}}.\sqrt{\sum_{i=1}^{n}(\textit{\^x}^{(i)})^{2}}}
\end{equation}
where $\mathbf{x}_{1},...,\mathbf{x}_{n}=\mathbf{X}\in\mathbb{R}^{N}$ and $\mathbf{\hat{x}}_{1},...,\mathbf{\hat{x}}_{n}=\mathbf{\hat{X}}\in\mathbb{R}^{N}$It is same as the closeness  between two non-zero vectors of an inner product space that measure the cosine angle between them. The reason behind using cosine is that the value of cosine will increase with decreasing value of angle which signifies more similarity. The vectors are length normalized after which they become vectors of length 1 and then the cosine calculation is simply the sum-product of vectors. 

\paragraph{Poisson Loss:}
The Poisson loss function\cite{losses} is a measure of deviation from the predicted distribution and the expected distribution. The loss function is computed by
\begin{equation}
\label{eq2}
\mathcal{L} = \frac{1}{n} \sum_{i=1}^{n} (\mathbf{\hat{x}}^{(i)} - \mathbf{x}^{(i)} . log(\mathbf{\hat{x}}^{(i)}))
\end{equation}
\fi

\subsection{Neural Network with DeepLDA}
\begin{figure*}
\includegraphics[scale = 0.3]{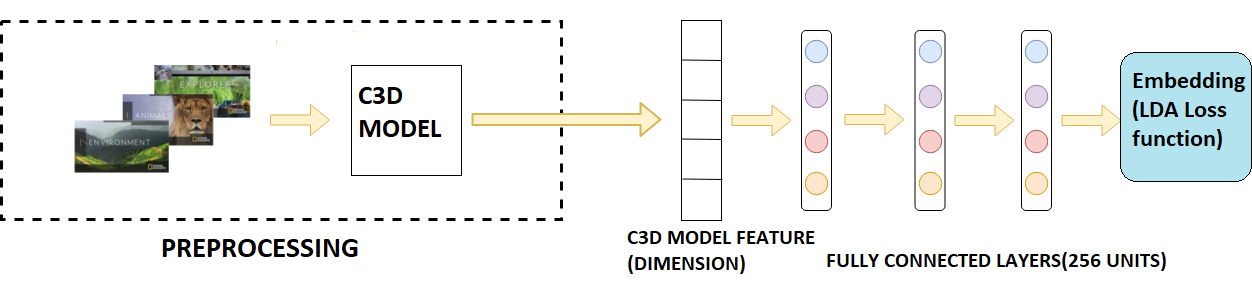}
\caption{DeepLDA based model's architecture}
\label{deepLDA}
\end{figure*}
\label{Deep Network with deepLDA objective function}
This architecture is composed of a deep neural network which uses the output of a C3D model i.e. the video-level features, as inputs. These are passed through a neural network containing two hidden fully connected layers, followed by a final output layer. A modified version of Linear Discriminant Analysis(LDA), is used as an objective function. Figure \ref{deepLDA} shows a high level view of this model.

\subsubsection{Model Architecture:}
\label{Model Architecture}
In this model, the video level feature set is used as the input to a set of three fully connected layers. The number of fully connected layers was decided empirically. Every fully connected layer involves, linear transformation of the input features, followed by a non-linear activation function. The first two layers, use the ReLU activation. The final layer serves as the output layer. As the C3D embeddings are limited to 512 dimensional feature vector, a smaller model with fewer parameters seem to give better results than a deep model with large number of hidden layers.

\subsubsection{Linear Discriminant Analysis Objective Function:}
\label{Linear Discriminant Analysis Objective Function}

\paragraph{Classic Linear Discriminant Analysis:}
Classic LDA, is a dimensionality reduction algorithm, which takes the higher dimensional data, and projects it onto a lower dimensional space such that the separation within data points of the same class is minimized and separation between data points of different classes is maximized\cite{fisherLDA}. Let $\mathbf{x}_{1},...,\mathbf{x}_{n}=\mathbf{X}\in\mathbb{R}^{Nxb}$ be a set of \emph{N} data points which belong to \emph{C} different classes. LDA tries to find a transformation matrix $\mathbf{A}\in\mathbb{R}^{lxb}$ which projects the input to a lower dimensional space \emph{L} where, \emph{l} = \emph{C}-1. The linear combination of features, obtained by multiplying the \textbf{X} with $\mathbf{A}^{T}$ has maximum separation in this lower dimensional subspace. The LDA formulation to find \textbf{A} is given as: 
\begin{equation}
\label{eqn1}
\underset{\mathbf{A}}{\operatorname{argmax}}
\frac{|\mathbf{A}\emph{S}_{b}\mathbf{A}^{T}|}{|\mathbf{A}\emph{S}_{w}\mathbf{A}^{T}|} 
\end{equation}
where, $\emph{S}_{b}$ is the between class scatter matrix and $\emph{S}_{w}$ is the within class scatter matrix. $\emph{S}_{t}$ denotes the overall scatter matrix. $\overline{\mathbf{X}}_{c}=\mathbf{X}_{c}-\mathbf{m}_{c}$ are the mean-centered observations for each class, the vector $\mathbf{m}_{c}$ represents the mean of each class similarly $\overline{\mathbf{X}}$ represents mean centered observation of the complete population. The formula for $\emph{S}_{t},\emph{S}_{w}$ and $\emph{S}_{b}$ is given as:
\begin{equation}
\label{eqn2}
\mathbf{S}_{c}= \frac{1}{\emph{N}_{c}-1}\mathbf{\overline{X}}_{c}^{T}\mathbf{\overline{X}}_{c}
\end{equation}
\begin{equation}
\label{eqn3}
\mathbf{S}_{w} = \frac{1}{C}\sum_{c}\mathbf{S}_{c}
\end{equation}
\begin{equation}
\label{eqn4}
\mathbf{S}_{t}= \frac{1}{\emph{N}-1}\mathbf{\overline{X}}^{T}\mathbf{\overline{X}}
\end{equation}
The value of transformation matrix \textbf{A} that maximizes Eq. \ref{eqn1}, maximizes the ratio of the `between class scatter' and `within class scatter'. Thereby, giving projections with low intraclass and high inter-class variance. To find optimum solution for Eq. \ref{eqn1}, we need to solve the eigenvalue problem of $\mathbf{S}_{b}\mathbf{e} = \mathbf{v}\mathbf{S}_{w}\mathbf{e}$, where \textbf{e} and \textbf{v} represent the eigenvector and their corresponding eigenvalues. The matrix \textbf{A} is the set of eigenvectors \textbf{e}. In the paper \cite{deepLDA}, authors have extrapolated this concept to introduce the DeepLDA objective function.

\begin{small}
\begin{table*}[t]
\caption{Track 1 and 2 - Results for the different architectures for TV shows and movies}
\label{results_shows}
\begin{tabular}{|l|l|l|llll|llll|} \hline
% RandomForest used C3d regression C3d+ (c3d+incept) + deep lda (c3d)
\textbf{Data}& \textbf{Data Subset} &\textbf{ Method} & &\textbf{Hit} &\textbf{@} & \textbf{k} &  &\textbf{Recall} &\textbf{@} &\textbf{k}\\   
 & & & k=5 & k=10 & k=20 &  k=30& k=50 & k=100& k=200 & k=300\\ \hline
%\textbf{Training Set} & -  & - & - & - & -\\

\multirow{8}{*}{\begin{sideways}
\textbf{TV Shows}
\end{sideways}} & \multirow{4}{*}{\textbf{Validation Set}} & Random Forest  & 0.004  & 0.016 & 0.063 & 0.112 & 0.017 & 0.034 & 0.062 & 0.084 \\
& & Regression(Cosine Loss)  &\textbf{0.319} &\textbf{0.388} &\textbf{0.485} &\textbf{0.541} &\textbf{0.171} &\textbf{0.229} &\textbf{0.289} &\textbf{0.323} \\ 
& & Regression(Poisson Loss) & 0.284 & 0.353& 0.432&0.483 &0.145&0.205&0.276&0.322\\
& & DeepLDA & 0.178& 0.234 &0.310 &0.354 &0.065&0.103&0.153&0.196\\
\cline{2-11}

%\textbf{Validation Set}& Random Forest Regression & 0.004  & 0.016 & 0.063 & 0.112 & 0.225\\
& \multirow{4}{*}{\textbf{Testing Set}}   & Random Forest & 0.047  & 0.093 & 0.178 &0.262 &0.031 &0.056 &0.093 &0.122\\
& & Regression(Cosine Loss)  &\textbf{0.226} &0.295 &0.38 &0.434 &\textbf{0.075 }&0.104 &0.137 &0.157\\ 
& & Regression(Poisson Loss) &0.217 &0.294 &0.381 &0.426 &0.076 &0.106 &0.139 &0.16\\
% Deep LDA filled 
& & DeepLDA & 0.22& \textbf{0.302}& \textbf{0.404}&\textbf{0.465}& 0.071& \textbf{0.117}&\textbf{0.18} &\textbf{0.224}\\
\hline

\multirow{8}{*}{\begin{sideways}
\textbf{Movies}
\end{sideways}} & \multirow{4}{*}{\textbf{Validation Set}} & Random Forest &0.004  & 0.011 & 0.037 & 0.063 &0.015 &0.025 &0.045 & 0.063\\

& & Regression(Cosine Loss) &\textbf{0.18} &\textbf{0.231} &\textbf{0.313} &\textbf{0.362} &0.099 &0.139 &0.192 &0.226\\
& & Regression(Poisson Loss) & 0.143& 0.191& 0.268 &0.318 &\textbf{0.114}&\textbf{0.166}&\textbf{0.246}&\textbf{0.313}\\

& & DeepLDA &0.106 &0.143 &0.192 &0.238 &0.065&0.091&0.129&0.158\\
\cline{2-11}
& \multirow{4}{*}{\textbf{Testing Set}}  & Random Forest & 0.055  & 0.109 & 0.172 & 0.216 & 0.035 &0.051 &0.077 &0.097\\

& & Regression(Cosine Loss)&0.108 &0.144 &0.212 &0.256 &0.044 &0.062 &0.084 &0.099\\

& & Regression(Poisson Loss) &0.107 &0.151 &0.212 &0.25 &0.044 &0.061 &0.082 &0.096\\
& & DeepLDA &\textbf{0.161} &\textbf{0.214} &\textbf{0.286}&\textbf{0.33} &\textbf{0.072}&\textbf{0.099}&\textbf{0.137}&\textbf{0.169}\\
\hline
\end{tabular}
\end{table*}
\end{small}

%\paragraph{DeepLDA optimization objective:}
%One of the issues encountered while solving the eigenvalue problem $\mathbf{S}_{b}\mathbf{e} = \mathbf{v}\mathbf{S}_{w}\mathbf{e}$, is the estimation of the within-scatter matrix $\mathbf{S}_{w}$, as it has been shown before\cite{problem1, problem2}, that within scatter matrix overemphasizes larger eigenvalues, while diminishing the lower eigenvalues. To overcome this issue, a multiple of the identity matrix is added to serve as a regularizing term that stabilizes the smaller eigenvalues\cite{solution1}.
%\begin{equation}
%\label{eqn5}
%\mathbf{S}_{b}\mathbf{e}_{i} = \mathbf{v}_{i}(\mathbf{S}_{w}+\lambda\mathbf{I})\mathbf{e}_{i}
%\end{equation}
In \cite{deepLDA}, the LDA function has been modified to be used as an objective function for deep neural networks by maximizing the individual eigenvalues. Each value gives the degree of separation along a particular eigenvector \emph{i.e.} higher eigenvalues correspond to a greater amount of separation. Maximizing these eigenvalues will help in obtaining an embedding in which classes are well separated. However, if we consider all the eigenvalues in the loss function, it can lead to a trivial solution in which only largest of the eigenvalue is maximized. To circumvent this problem, we optimize only on smallest \emph{m} eigenvalues out of total set of eigenvalues. This can be done by using only those \emph{m} eigenvalues that are less than a certain predetermined threshold. Therefore, the formulation becomes:
\begin{equation}
\label{eqn6}
\underset{\mathbf{\theta}}{\operatorname{argmax}}\frac{1}{\emph{k}}\sum_{i=1}^{k}\emph{v}_{i} \:with\: \{\emph{v}_{1},...,\emph{v}_{k}\}= \{\emph{v}_{j}|\emph{v}_{j}< \min{\emph{v}_{j}|\emph{v}_{j}}+ \epsilon\}
\end{equation}
This loss is back-propagated through the network. In our model, each row in the \textbf{relevance list} specifies a batch. All the movies/shows that come in the relevance list of a particular show/movie are given one class and the others which are not in list are assigned a separate class. Then, we train on the loss using ADAM optimization\cite{adam}.

\section{Evaluation}
\label{evaluation_of_the_model}

\subsection{Evaluation Metric}
For the challenge, we use two evaluations measures\cite{cbvrp-acmmm-2018} -

\begin{enumerate}
\item Recall@K - For top $K$ predictions the metric \textit{recall@K} is defined as - 
\begin{equation}
	recall@K = \frac{|o^{r} \cap o^{\overline{r}}|}{o^{r}}
\end{equation}
where $o^{r}$ is relevance list and and $o^{\overline{r}}$ predicted top K relevant shows/movies for each item.
\item Hit@k - If \textit{recall@K} > 0 for a test case then \textit{hit@K} has value 1 or else value 0 for that test case.
\end{enumerate}
The average values of recall and hit is taken on all test cases for evaluation purposes \cite{cbvrp-acmmm-2018}.

These evaluation metrics have been used in several other works such as \cite{li2017study,deldjoo2015toward, sedhain2014social}. 

\subsection{Results}

Table \ref{results_shows} contains the results obtained from all the different architectural models considered for both TV shows as well as movies obtained on validation set and training set after training the models. For various values of $K$, the results are presented in the table.

The DeepLDA model clearly gives the best results on these metrics for both TV shows and movies on the testing set. The competition used \emph{recall@100} and \emph{hit@30} for the evaluation of results, thus using that the analysis of results is presented below.
%while on the validation set, regression model with cosine loss performs the best for TV shows. For movies, regression model with cosine loss gives higher hit rate while with poisson loss gives higher recall rate.

For TV Shows, considering the hit values on the test data, the DeepLDA model performs better than the Random Forest model by 77.4\%, Regression model with cosine loss by 7.1\% and Regression model with poisson loss by 9.1\%. For the metric recall, the DeepLDA model performs better than the Random Forest model by a full 108\%, Regression model with cosine loss by 12.5\% and Regression model with poisson loss by 10.3\%. 

For movies, considering the hit values on the test data, the DeepLDA model performs better than the Random Forest model by 52.7\%, Regression model with cosine loss by 28.9\% and Regression model with poisson loss by 32\%. Taking into accoutn recall values, the DeepLDA model performs better than the Random Forest model by 94.1\%, Regression model with cosine loss by 59.6\% and Regression model with poisson loss by 62.2\%.

\section{Conclusions and Future Work}
\label{conclusion}
The paper presented several models on which extensive experiments were done to provide content based recommendations to users. This solution provided a remarkable solution for the `cold-start' problem faced by video recommendation engines like Hulu. The relevance between two videos is computed using different types of losses, regression techniques and linear discriminant analysis.

In the future, the authors aim to train models using hybrid models generated after employing explicit as well as implicit features.  These models have the potential of improving the results for video recommendation significantly. Additionally, they also look forward to extend the current models for addressing the pernicious cold start problem holistically by taking into account not just cold videos but cold-start user profiles too. These approaches could also help in ameliorating the accuracy of the models considered and make it more robust.

%\end{document}  % This is where a 'short' article might terminate